\documentclass[twoside,11pt]{article}

\usepackage{automl2020}
\usepackage{multicol}
\usepackage{amsmath}
\usepackage{floatrow}
\usepackage{mathtools} 
\newfloatcommand{capbtabbox}{table}[][\FBwidth]
\newfloatcommand{capbfigbox}{figure}[][\FBwidth]
\usepackage[ruled,vlined]{algorithm2e}
\usepackage{multirow}
\usepackage{array}
\usepackage{subcaption}

\graphicspath{ {./figures/} }
\usepackage{comment}
\usepackage{graphicx}

\newcommand{\splitatcommas}[1]{\begingroup\lccode`~=`, \lowercase{\endgroup
    \edef~{\mathchar\the\mathcode`, \penalty0 \noexpand\hspace{0pt plus 1em}}%
  }\mathcode`,="8000 #1%
  }

\jmlrheading{S. Ackerman, P. Dube, and E. Farchi}

\ShortHeadings{Sequential drift detection in deep learning}{Ackerman, Dube, and Farchi}
\firstpageno{1}

\begin{document}

\title{Sequential drift detection in deep learning classifiers}

\author{\name Samuel Ackerman \email 
       samuel.ackerman@ibm.com \\
       \addr IBM Research, Israel
       \AND
       \name Parijat Dube \email pdube@us.ibm.com \\
       \addr IBM Research, USA
         \AND
       \name Eitan Farchi \email 
       farchi@il.ibm.com \\
       \addr IBM Research, Israel
     }
\maketitle

\begin{abstract}


We utilize neural network embeddings to detect data drift by formulating the drift detection within an appropriate sequential decision framework.   This enables control of the false alarm rate although the statistical tests are repeatedly applied.   Since change detection algorithms naturally face a tradeoff between avoiding false alarms and quick correct detection, we introduce a loss function which evaluates an algorithm's ability to balance these two concerns, and we use it in a series of experiments.




\end{abstract}

\section{Introduction}\label{sec:intro}

In an earlier work \citep{farchi_dube20} we presented a framework for automatic detecting of drift in images received by a neural network-based classifier.  In these experiments, the stream of data objects observed changed from one class to another (e.g., from images of animals to flowers) by being contaminated by images of the second class at various rates.  By using an embedding of the second-to-last layer of the network---that is, the layer of neurons before the class output---we were able to robustly capture information of the received images, and detect the data drift when the nature of the inputs changed.  This technique is successful in transferring complex drift detection problems to a univariate problem.

Embedding of the inner layers of the neural network has been previously applied to "explain" a decision made by the network on a specific data point (see \cite{DBLP:journals/corr/abs-1708-08296}).   Given a data point $x = (x_i)_{i = 1, \ldots, n}$, the network classifies $x$ as some object.   Typically that will include some output $f(x)$ of the last neuron that is then threshold to make the final classification decision.   Using an explainability scheme, contribution at each neuron to the output $f(x)$ is obtained.  This is propagated back through the layers of the network resulting in an explanation of the decision at each vector component $x_i$ of the data point.  Our work borrow this concept of associating embedding to the inner lawyers of the network but applies it to design a statistics that will identify drift in the ML model performance.  

In the experiments in \citet{farchi_dube20}, we repeatedly applied the non-parametric Mann-Whitney (MW) Test \citep{mann47,nachar08} to detect change in the distribution of the embedding values.  Here, we compare our earlier results with a different testing framework, which is \textit{sequential} in that it explicitly considers the fact that the data arrive over time and that repeated testing is performed.  To that end we use the CPM sequential model \citep{AdamsRoss_cpm_2015}.  A false alarm, also known as a Type-1 error, occurs if we mistakenly decide that drift has occurred when it hasn't (yet).  The purpose of the sequential methods that we discuss is to appropriately ensure that the probability of a false alarm happening is kept below some low known value called $\alpha$ (often 0.05).  If done, this gives us a statistically-based assurance that our decision that drift has occurred is very likely correct.

We introduce a new loss function to score the MW and sequential methods' abilities to detect drift quickly and correctly (avoiding false alarms) over various datasets and drift contamination scenarios. 



\section{Methodology}\label{sec:methodology}
\subsection{Problem definition}
\label{sssec:problem_definition}

Below, we provide a summary of our experimental set-up from \citet{farchi_dube20}; the reader is directed to the paper for complete details.

A neural network classifier $M$ is first trained on a baseline set of images $T$ that belong to a given class (e.g., `animals').  A detection algorithm $\mathcal{A}$ will monitor the stream of images in sequence.  After a fixed (non-drift) period of receiving new images from the same class as the training set $T$, we intersperse the image stream with images of a different class (e.g., `plants') and see if $\mathcal{A}$ can detect the change in inputs. 

As noted by \citet{MRACH_2012}, there is substantial confusion in the use of terms such as `concept' and `data' drift in settings where a concept or target $Y$ is modeled based on data $X$; therefore, we adopt their notation in defining our problem.  We say that the problem of predicting an image class label $Y$ (e.g., `animal') based on image input $X$ (e.g., a pixel vector) is actually a $Y\rightarrow X$ problem since saying an image is of a `dog', say, determines what the object in the image ($X$) will look like; this is much like the example they cite of a disease definition ($Y$) determining the symptoms.  Here, $\textrm{Pr}(X\mid Y)$---that is, for instance, ``what a typical image looks like if it is an animal"--- remains unchanged throughout, only $\textrm{Pr}(Y)$ changes, since we deliberately change the class sampling distribution by inserting new class images.  They term this a `prior probability shift' problem.  However, we note that our experiments, which model changes in the data $\mathbf{X}$ itself (though focused on detecting changes in $\textrm{Pr}(\mathbf{X})$ due to changes in $\textrm{Pr}(Y)$), is very general and can detect various drift types.

In addition, the literature on monitoring data change has long recognized (e.g., \citet{KR_1998}) that the rate of change over time between drift contamination distributions (the proportion of new class images) affects the success of detection.  We therefore experiment with `linear' (gradual, or `drift') and `step' (abrupt, or `shift') changes (see Sections~\ref{sec:loss_function} and \ref{sec:eval}).

\subsection{Image feature representation}
\label{sssec:image_feature_representation} 

For our detection mechanism $\mathcal{A}$, we transform the problem of detecting image class prior probability shift from the high-dimensional (and image domain-specific) setting to a univariate numeric problem.  Such an approach is common (e.g., \citet{KR_1998}, \citet{L_1999}, \citet{LMD_2011}, \citet{orna19}), for instance, monitoring changes in model confidence, estimated accuracy, or other outputs, since univariate metrics are more easily analyzed by methods such as density estimates and statistical tests. Unlike many of the works cited above, our experiments are aimed at detecting changes in the data $\mathbf{X}$ or class $Y$ indirectly through monitoring a univariate divergence representation of the changes in the data, rather than at detecting changes in the predictive performance of a model through monitoring metrics such as accuracy.  Our approach is model-specific in that it operates on a neural network, but the purpose of this is to use the internal model representation (see below) of the data to monitor data changes, since neural networks have powerful data-representation capability and are widely-used and generally-applicable.

In our experiments, new images are received in batches of fixed size $b_s$; overall, we receive $N=120$ batches, indexed by $t$. For each image, let $f_n$ be the average feature vector extracted from neuron layer $n$ (immediately before the output) of network $M$.  For a given batch of images, let $\hat{f}_n^t$ be the element-wise vector average of their respective vectors $f_n$ in batch $t$, to reduce sensitivity to the particular order of images. Let $\hat{f}_n^T$ be the respective average vector for the training set $T$.  Let $\mathcal{D}_{T,t;n}$ be an calculated divergence measure between the feature vectors for $T$ and batch $t$.

There are other examples in the literature of using neural network embeddings for non-standard tasks (that is, for purposes other than simply outputting a prediction).  For instance \citep{embedCluster} use a neural network to embed inputs and perform clustering on the embedded rather than original space.

The images are not observed directly by our detector $\mathcal{A}$, only the divergences \splitatcommas{$\mathcal{D}_{T,1;n},\mathcal{D}_{T,2;n},\dots,\mathcal{D}_{T,120;n}$} for each batch $t$.\footnote{We note that this differs from our setup in \citet{farchi_dube20} in which the observed divergences were $\mathcal{D}_{t-1,t;n}$ between consecutive batches $t-1,t$ instead of between each batch and the fixed baseline set $T$; our new approach should allow easier identification of changes.} Between batch $t=t_s$ to $t_e$ (start and end of drift), images from another class (the prior probability shift) are gradually mixed in with the first class, to be detected only indirectly through changes in the observed divergences $x_t=\mathcal{D}_{T,t;n}$.  We note that this has two prime benefits, both of which assume that $M$ well-captures relevant image characteristics.  Firstly, extracting feature representations generalizes the problem away from the specific domain of images and eliminates the need to construct image-specific metrics.  Secondly, although a human can typically easily distinguish between `animals' and `plants', for instance, not only would monitoring many images manually be cumbersome, but also a human can miss subtle changes in the data that may be important (that is, in $\textrm{Pr}(X\mid Y)$), such as changes in the image resolution or sub-types of `animals'.

We now outline the drift detection algorithm from \citet{farchi_dube20}.  We set a sliding window of size $\beta=40$, which is split into two halves $D_1$ and $D_2$, each of size $\beta/2$, where $D_1$ is first.  Beginning at time $t=40$ (the earliest to have a full $\beta$ points), a non-parametric
MW test is performed, testing whether the samples $D_1=\{x_{t-\beta+1},\dots,x_{t-\beta/2}\}$ and $D_2=\{x_{t-\beta/2+1},\dots,x_t\}$ differ in distribution.  If so, drift is declared at time $t$ to have happened at some point in the past $\beta$ points.  It is expected that the divergence $\mathcal{D}_{T,t;n}$ when $t<t_s$ (before drift) should be low (since they are of the same class as the training set) and should be higher when $t\geq t_s$ after drift is introduced.  See Appendix~\ref{appendix_mw} (Algorithm~\ref{alg:mw}) for the full algorithm.

\section{Sequential Error Control}
\label{sec:sequential}

\subsection{Sequential statistical testing}
\label{sssec:approach}

The MW Test setup from \citet{farchi_dube20} was a non-sequential test.  That is, the test was applied repeatedly on each sliding window, and the p-value was used to count one instance of a detected difference (five instances were used to declare drift) without considering the fact that the test was conducted repeatedly.  However, it is well-known that repeated statistical testing without appropriate adjustment will cause unacceptable rates of false alarm (Type-1 errors). In the drift setting, the null hypothesis ($H_0$) will be that drift has not occurred, while the alternative ($H_A$) is that it has.  Consider a test for identifying drift, which has decision threshold $\alpha$ (e.g., 0.05).  If applied on a single window of non-drifted data ($H_0$), this test will raise a false alarm (falsely say there is drift) with probability $\alpha$.  This $\alpha$ is a factor known by the user.  However, say the test is applied to $w$ independent windows of non-drifted data, and that drift is declared on \textit{any} of the $w$ batches (that is, its p-value is $<\alpha$).  The probability of the correct decision here (that is, of none of the windows falsely indicating drift) is now $(1-\alpha)^w$, rather than the higher $1-\alpha$ for a single test.  Thus, a test naively applied multiple times without proper adjustment will not give the expected $\alpha$-level statistical guarantee.

The rationale for making repeated drift tests in this setting is that we want to detect drift as soon as possible, and hence have to conduct the test at multiple time points without waiting to observe all the data; in many cases, the data may be an `infinite' stream without a predetermined sample size.
The problem of false alarms under multiple tests illustrated above is compounded when the data windows are not independent, as in the case  
where they overlap when we want to examine overlapping windows of historical values.  Such is the case in our Algorithm 1, in which successive windows of size $\beta$ overlapped.  In order to use the entire history to detect drift, while controlling false alarms, an additional adjustment is needed.  See Appendix \ref{sssec:peeking_problem}.

\subsection{Change Point Models (CPMs)}
\label{subsec:cpm}

As noted by \cite{RATH_2012} in presenting their ECDD method for detecting concept drift in Bernoulli-distributed variables (e.g., binary indicator of correct classification), many drift detection methods suffer a weakness in that they cannot properly demonstrate that the false alarm probability is controlled in reality under a wide variety of settings.
That is, for instance, to use their example, since the rate of positive instances (instances of drift) in the data stream is unknown, if the creators of a given method experimentally demonstrate that, say, their method makes one false alarm detection every 100 observations, this may be too bad of a result and would lead us to regard the positive detections of such a method as false positives.  One false positive every 5,000 observations rather than 100 may be needed to demonstrate that a positive decision of drift is to be trusted. 

It seems that many methods of drift/shift detection may thus use data streams that are too short relative to what is likely to be encountered in a realistic scenario; this is particularly true if a method must make a single positive decision and be evaluated based on it (i.e. `all-or-nothing' scoring) rather than being allowed to make more than one false positive and being evaluated on the average rate.
A method with an average run length ($\textrm{ARL}_0$, the average number of observations between false positives) of 100 may simply be inadequate if most of its decisions turn out to be false positives and a significant effort (e.g., re-training a model or re-evaluating the data) need be taken each time it makes a decision of drift.

We give brief examples of how (concept) drift detection methods have typically demonstrated success.  \cite{LMD_2011} calculate a signal value on the data, and count how often the signal value is observed to be a given number of standard deviations above the mean, calculated historically.  They try declaring drift by trying several rules for how many times this has to happen ($x$) in the past ($y$) observations (`Western Electric Rules'); we note this is a similar strategy to our work in \cite{farchi_dube20}.  \cite{SK_2017} give an intriguing method to detect ML model potential failure by monitoring a model's ``regions of uncertainty" through its ``marginal density".  The thresholds for change detection are based on the  empirical mean and standard deviations of their metric under K-fold cross validation.  While the experimental results seem adequate, it does not seem that they attempt to control the false positive rates either sequentially or using the criteria of \cite{RATH_2012}.

The sequential statistical technique we adopt to deal with these issues is the Change Point Model (CPM) developed by \citet{AdamsRoss_cpm_2015} and implemented as the \texttt{cpm} package \citep{cpmR2015} for \texttt{R} software. The CPM allows us to conduct repeated backwards-looking drift detection while controlling false alarm probability (Type-1 error) for a user-desired value of $\textrm{ARL}_0$; this method also has theoretic statistical guarantees on correctness, not just limited experimental results, as we summarize below.  See Algorithm~\ref{alg:cpm} in~\ref{appendix_cpm} for an outline of the CPM for input data $x_1,x_2,\dots$ observed sequentially.  We note also that two of the authors of the CPM (Adams and Ross) were authors of the ECDD \citep{RATH_2012} discussed above.

\begin{figure}[h]
    \centering
    \includegraphics[scale=0.47]{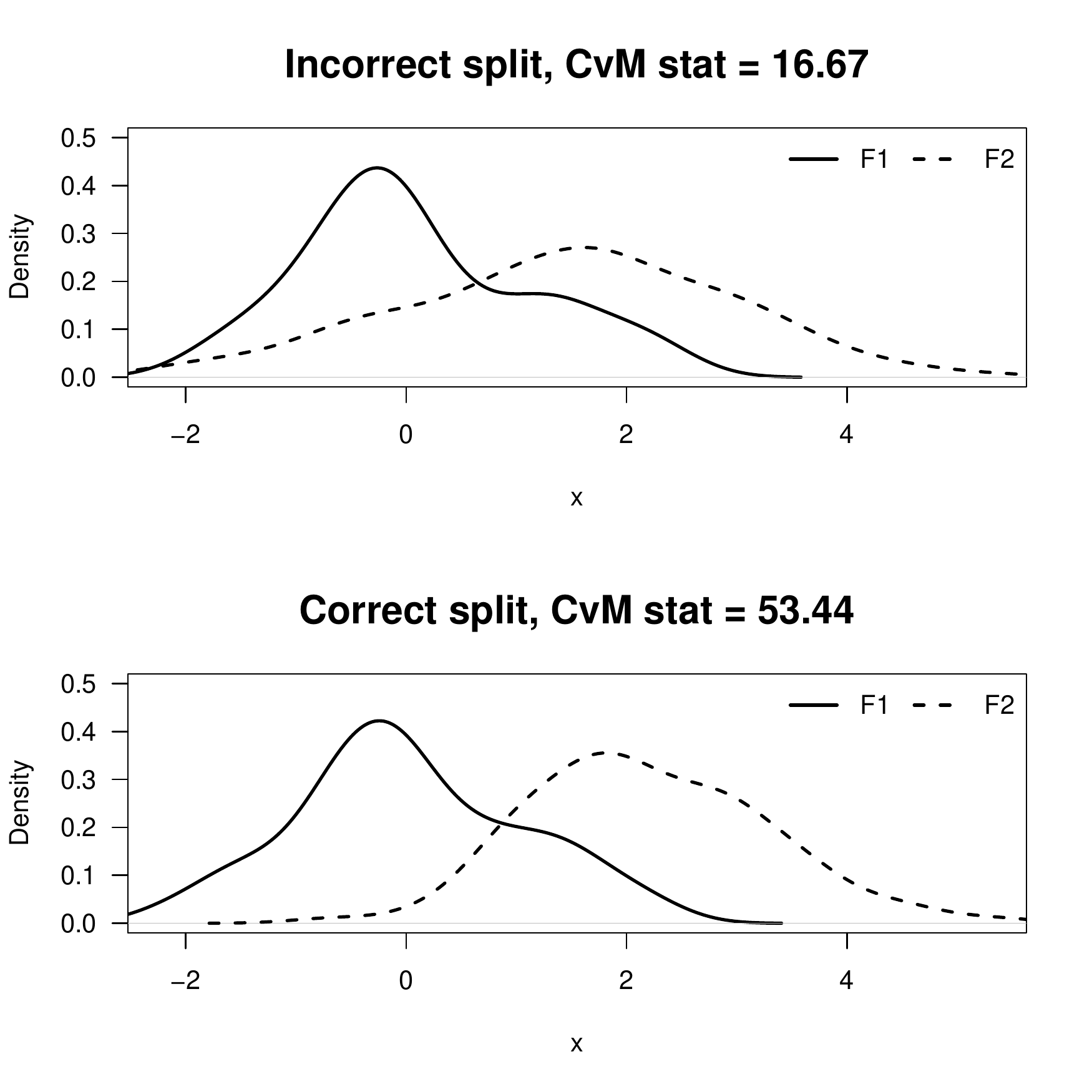}
    \caption{\label{fig.cvm_split_toy}Illustration of value of Cramer-von-Mises (CVM) statistic value at the incorrect and correct split.  The statistic will tend to be maximized if performed at the correct split.}
\end{figure}

An overview of the CPM follows.  Conceptually, as shown in Algorithm~\ref{alg:cpm}, after a large enough initial stabilization period, $B$, at each time $t\geq B$, each possible changepoint (time at which drift began) $k=2,\dots,t-1$ is considered.  For each $k$, the data is split into before/after $k$ samples $\textbf{X}_0$ and $\textbf{X}_1$ of at least size 2.  If $k$ is a changepoint, $\textbf{X}_0\sim F_0$ and $\textbf{X}_1\sim F_1$ should represent two different distributions $F_0$ and $F_1$. The function $\text{Diff}(\cdot,\cdot)$ is a (non-parametric) two-sample test which tests if $F_0$ and $F_1$ differ, and returns a statistic $W_{k,t}$, which are then normalized; several tests, such as Student, Cramer-von-Mises, and Kolmogorov-Smirnov, are provided.  $\hat{\tau}$ is the $k$ which maximizes $W_{k,t}$.  If $\hat{\tau}$ is around the true changepoint (if it occurred), the statistic $W_t=W_{\hat{\tau},t}$ should be large, as illustrated in Figure~\ref{fig.cvm_split_toy}.  If $W_t>h_t$, where $h_t$ are critical values based on simulations of these statistics, the changepoint $\hat{\tau}$ is a significant enough split that the sample after, $\textbf{X}_1$ appears to represent a different distribution, and thus `drift'.

\begin{figure}[h]
    \centering
    \includegraphics[scale=0.47]{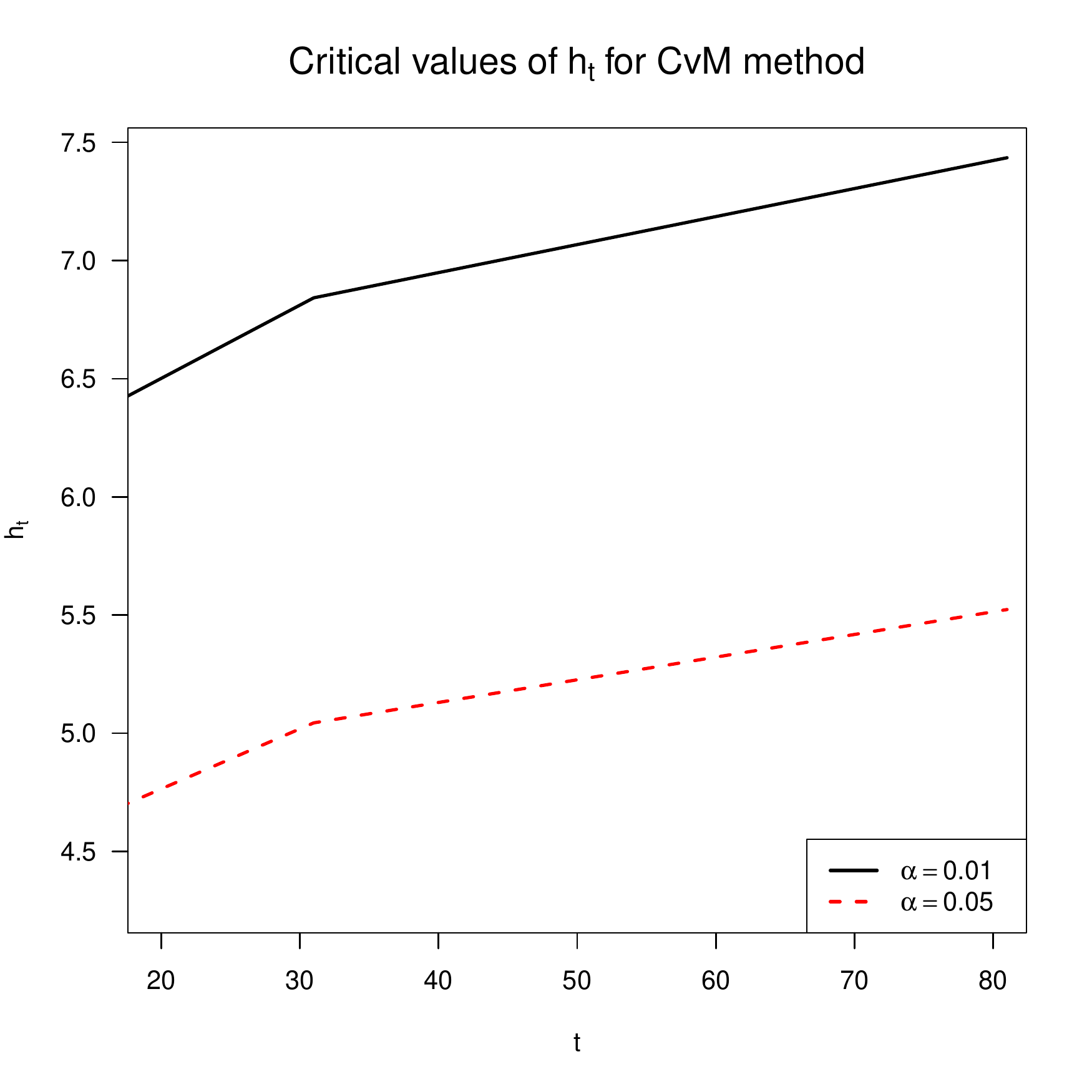}
    \caption{\label{fig.h_t}Critical values $h_t,\: t=20,\dots,80$ for the Cramer-von-Mises CPM for $\alpha=0.01, \:0.05$.  For a lower $\alpha$ (higher significance), the critical values are higher.  Also, for each $\alpha$, $h_t$ increase with $t$.}
\end{figure}

Crucially, the critical values $h_t$ (which depend on the two-sample test used) increase with $t$ (see Figure~\ref{fig.h_t}) so that the following inequalities are satisfied: 
\begin{itemize}
    \item $\text{Pr}(W_1>h_1\mid \text{no change})=\alpha$
    \item $\text{Pr}(W_t>h_t\mid \text{no change by $t$ and } W_{t-1}\leq h_{t-1},\dots,W_1\leq h_1)=\alpha, \:\forall t>1$
\end{itemize}
Together, these imply that at any time $t$, assuming we have not declared drift yet (otherwise we would not be observing $x_t$), if there has not been drift yet ($H_0$ true), the probability of a false alarm ($W_t>h_t$ mistakenly) is always $\alpha$.  This gives us the desired false alarm control.  The reason we choose the CPM (out of many possible  sequential methods) is due to this false alarm control guarantee, which applies at each time $t$; this is precisely what other methods typically do not do, they typically rely on average results without properly taking into account the length of time observed.    Furthermore, the tests are non-parametric, meaning they do not assume that $x_t$ follow a particular distribution.

Thus, our sequential modification to the non-sequential MW algorithm is to apply the CPM (with a test such as Cramer von-Mises) on a sequence of calculated divergences ${\cal D}_{T,t;n}$. Note that drift does not mean that  ${\cal D}_{T,t;n}$ change from being near 0 to becoming large, but rather that the distribution of typical observed divergences has changed, which indicates the underlying data have changed. 

The CPM modification of the non-sequential Algorithm~\ref{alg:mw} is simply to apply Algorithm~\ref{alg:cpm}, where again $x_t=\mathcal{D}_{T,t;n}$, the observed divergence; we use the Cramer-von-Mises version of the CPM for generality.  In addition to the statistical false alarm control, there are several differences between the two algorithms:
\begin{itemize}
    \item At each $t$, the CPM examines the entire past history, and not just the past $\beta$ points as in Algorithm~\ref{alg:mw}. 
    \item The CPM considers all possible split points $k=2,\dots,t-1$ and finds the separating point $\hat{\tau}$.  This has two advantages. First of all, it can pinpoint the likely time $\hat{\tau}$ of drift, which the MW cannot.  Also, the ability to split the data into two probably unequal-size partitions before and after $\tau$, rather than fixing the two samples at an equal $\beta/2$, means that the drift identification may also be more precise than MW.  The second set $D_2$ will likely contain a mix of drift and non-drift class, since it must be of size $\beta/2$, while the CPM can ideally split the history into all non-drift (before $\hat{\tau}$) and all-drift (after).  As illustrated in Figure~\ref{fig.cvm_split_toy}, the most significant split statistic should occur at the true split point, and should be less significant if the samples must be of fixed size $\beta/2$.  %
\end{itemize}

\section{Loss function}
\label{sec:loss_function}

We have mentioned that the CPM provides a correct way of sequentially controlling the false alarm rate.  While guaranteeing low false alarm probability is important, so is detecting drift as quickly as possible.  We thus propose a penalty function which scores a drift detection algorithm across our experiment datasets, while assigning different importance scores to achieving these two goals. In \cite{farchi_dube20}, we experimented with simulating gradual vs linear drift contamination, and we also want drift to be detected more quickly the higher the drift class contamination is.

\begin{figure}[!htb]
    \centering
    \includegraphics[scale=0.25]{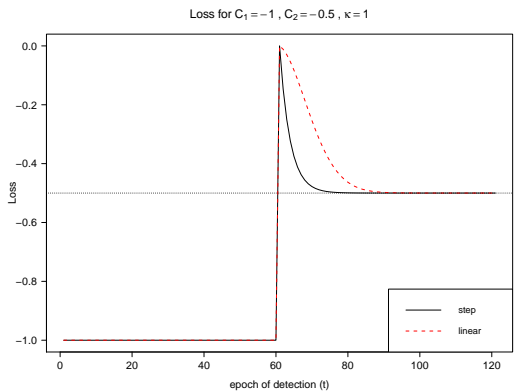}
    \caption{\label{fig.loss_function}Penalty function $g(t_s=60, t_d, \textbf{p})$ value for detection at epoch $t_d=0,\dots,120$ and two different contamination vectors $\textbf{p}$  From $t_d=60$ to 120, the penalty $g$ is decreasing.  Linear drift loss decreases more slowly than step drift because the drift contamination is gradual rather than sudden between times $t_s$ and $t_e=80$.}
\end{figure}

In our experiments, divergences $x_t=\mathcal{D}_{T,t;n},\:t=1,\dots,120$ are observed (see Section~\ref{sssec:image_feature_representation}), where $x_t$ is calculated from the $t^{\textrm{th}}$ batch of images (not observed by the user). Drift was introduced beginning at batch $t_s=60$ and reached full saturation by batch $t_e=80$.\footnote{Note: that according to CPM notation, the changepoint is actually $t_s-1$, the last index before drift.}  Hence, a detection at $t_d$ is a false alarm if $t_d<t_s$, and otherwise is a correct decision with delay $t_d-t_s$, where $t_d-t_s+1$ drift windows have been observed.  If no detection is made within the allotted time limit, let $t_d=\infty$.  Let $\textbf{p}=\{p_1,\dots, p_{n-t_s+1}\}, \: 0<p_j\leq 1$ be the drift contamination proportion in the 61 ($=n-t_s+1$) time windows beginning with the introduction of drift at $t_s$. Define $C_1,\: C_2\leq 0$ be penalties for a false alarm at any point and the greatest penalty for a (late) true drift detection.  Typically, we might set $C_1\leq C_2$ because a late, correct decision is better than raising a false alarm.  Define our penalty as
\[
g(t_s,t_d,\textbf{p})=
\begin{dcases}
C_1 \quad\text{   if } t_d < t_s\: (\text{false alarm})\\
C_2 -\left(\frac{\kappa C_2}{\prod_{j=1}^{t_d-t_s+1}(1+p_j)^{\frac{(t_d-t_s+1)-j}{t_d-t_s+1}}}\right) \\ \quad\quad\text{ if } t_s\leq t_d <\infty\: (\text{correct decision})\\
C_2 \quad\text{   if } t_d=\infty \: (\text{not detected})
\end{dcases}
\]
For a detection at $t_d \geq t_s$, the drift fraction $p_j$ observed at time $t_s\leq j \leq t_d$ has been observed for $t_d-j+1$ time points (including at $t_d$).  The penalty function $g$ compounds the drift contamination $p_j$ for each $j$ for the number of time points the drift detector has had a chance to observe it.  For instance, the best case is to detect the drift immediately at first introduction ($t_d=t_s$). Thus, the penalty is $g(t_s,t_s,\textbf{p}=\{p_1\})=C_2 - \frac{\kappa C_2}{(1+p_1)^\frac{(0+1)-1}{0+1}}=C_2-\frac{\kappa C_2}{1}=(1-\kappa)C_2$, which is the minimum penalty; if $\kappa=1$ (the default value) in this case, the penalty is 0 since it is the best result.  Any missed detection receives a penalty of $C_2$.  In general, any detection $t_d\geq t_s$, parameter $0\leq\kappa\leq 1$ controls the rate of decay of the penalty from $0$ to $C_2$ with respect to $t$, given $\textbf{p}$.
Figure~\ref{fig.loss_function} illustrates this penalty for $C_1=-1$ and $C_2=-0.5$.  Since the constants factor out of the equations, all that matters for comparing algorithms is the ratio between them.

In our experiments, we allow the method to make multiple false alarms, but only (up to) one true detection.  Let $\textbf{t}$ be a vector of drift detection times $\{t_d\}$, where only the first correct decision $t_d\geq t_s$ is included, or $\textbf{t}=\{\infty\}$ if no detection is made. The overall loss  $L(t_s,\textbf{t},\textbf{p})$ is thus 
$\sum_{t_d\in \textbf{t}}g(t_s,t_d,\textbf{p})$.

\section{Experimental results}
\label{sec:eval}

In our experiments, we simulate drift between images from different datasets (see Table 2 of~\cite{farchi_dube20} for dataset details) in 15 different combinations.  In each, the first $t_s-1=59$ batches were all from the first class.  The batch size in all the experiments is 16 images.  We used two different drift scenarios (which determine the drift contamination vector $\textbf{p}$):

\begin{itemize}
    \item \textbf{Step drift}: drift is introduced suddenly and fully beginning at $t_s$, that is, all batches from $t_s$ after are from the second class ($\textbf{p}=[1,1,\dots,1]$).
    \item \textbf{Linear drift}: second class is introduced in constant-increase gradual proportions $p_t=\frac{t-t_s+1}{t_e-t_s+1},\: t_s\leq t\leq t_e$, and $p_t=1,\: t>t_e$. 
\end{itemize}

For each drift scenario, we also evaluate our algorithm where the observed divergence $\mathcal{D}_{T,t;n}$ is either Kullback-Leibler Divergence (KLD) or cosine distance.  Note that since KLD is defined over probability distribution, we need to appropriately normalize the average feature vectors. The results for detection using either the sequential CPM (Subsection~\ref{subsec:cpm}) and non-sequential MW are compared using the loss function $L(t_s,\textbf{t},\textbf{p})$ (Section~\ref{sec:loss_function}).

Each algorithm returns a vector $\textbf{t}$ of detection times $t_d$, where $\textbf{t}=\{\infty\}$ if no detection is made. In addition to the loss function, we can calculate the overall false alarm rate as follows:  
\[
\text{false alarm rate ($\theta$)} \: = \: \frac{\mbox{no. of false alarms}}{\mbox{no. of false alarms} + I (t_s \leq t_d<\infty)}, 
\]
where $I(t_s \leq t_d<\infty)=1$ for correct detection and is 0 otherwise. Note that if there is no detection (false or not) then $\theta$ is undefined. Assume the time we have identified the drift is $t_d$, then detection delay ($\Delta$) is $t_d - t_s+1$

\begin{figure}[!tbp]
   \centering
   \begin{subfigure}{0.45\textwidth}
   \includegraphics[scale=0.45]{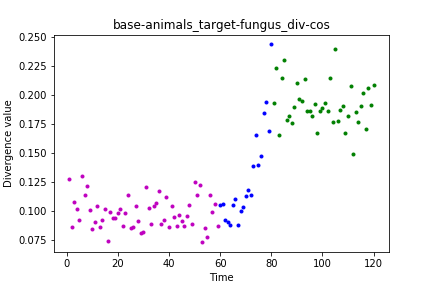}
   \caption{cosine}
   \end{subfigure} 
   \hfill
  \begin{subfigure}{0.45\textwidth}
   \includegraphics[scale=0.45]{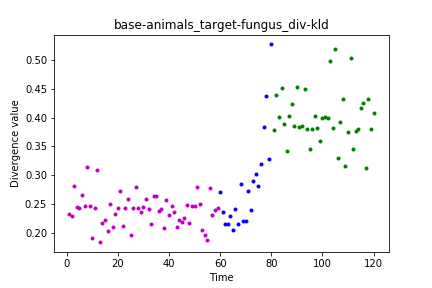}
 \caption{kld}
 \end{subfigure}
\caption{Divergence distribution over the experiment interval for an example scenario using cosine distance and KLD. Drift starts at $t=60$ and ends at $t=80$. The value of divergence is close to 0 for $t < t_s$ (magenta), has an increasing trend for $t_s \leq t \leq t_e$ (blue) and remains high for $t > t_e$ (green). \label{fig:div-distribution-linear}}%
\end{figure}

Here, we show results for the scenario of linear drift only; the step drift results are shown in the appendix in Subsection~\ref{subsec:step_drift}.
%
%
We first plot the distribution of divergence values during the entire time period of observation. 
This will visually show us how the divergence value evolve in three time intervals, before the start of drift period ($t < t_s$), during the drift period ($t_s \leq t\leq t_e$, and after the end of drift period ($t>t_e$). 
Figure~\ref{fig:div-distribution-linear} shows the divergence distribution calculated using cosine distance for two example scenarios. 

\begin{figure}[htp]
   \centering
   \begin{subfigure}{0.45\textwidth}
   \includegraphics[scale=0.45]{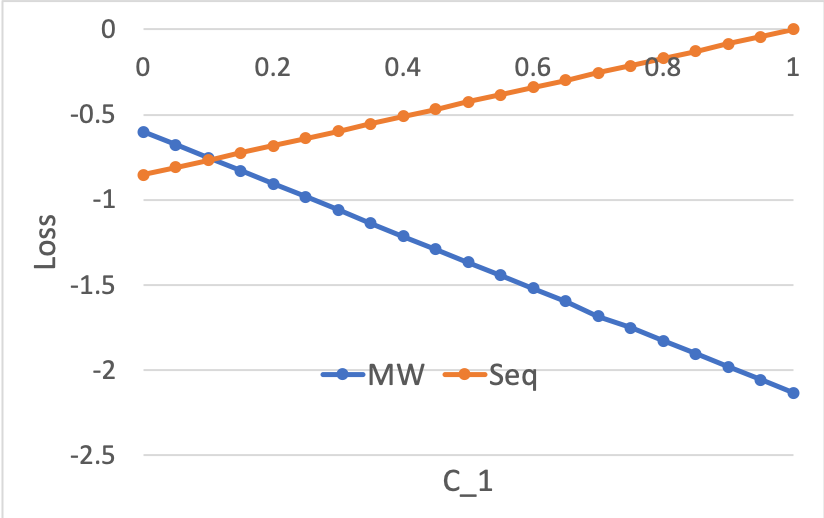}
   \caption{linear}
   \end{subfigure}
   \hfill
   \begin{subfigure}{0.45\textwidth}
 \includegraphics[scale=0.45]{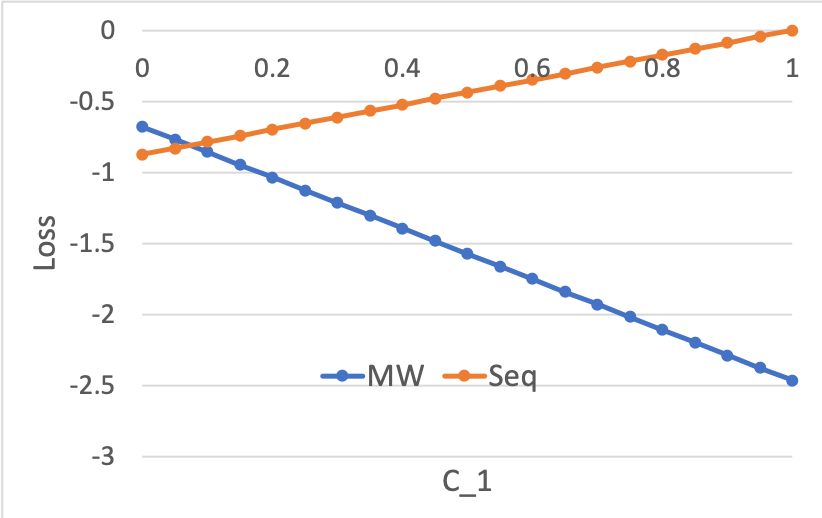}
 \caption{step}
 \end{subfigure}
\caption{The value of loss with MW and Seq as $C_1$ changes from 0 to 1 and $C_2=1-C_1$. While loss with Seq is mostly lower, for both linear and step drift scenarios there are small regions (when $C_1$ is very small) where loss with MW is lower compared to Seq. \label{fig:loss-linear-step}}%
\end{figure}

We next compare the sequential algorithm with the algorithm in \citet{farchi_dube20} using the loss function in Section~\ref{sec:loss_function}. Table~\ref{tab.experiments.linear} compares the two algorithms using KL and cosine divergence.  Compared to sequential, the MW-based detection algorithm has a lower detection delay but higher false alarm rate. The overall loss with $C_1=C_2=-0.5$ is higher with MW compared to sequential. However, depending on the ratio between $C_1$ and $C_2$, one can find regions where one algorithm is preferable over the other as shown in Figure~\ref{fig:loss-linear-step}.

\begin{table}[h]
\centering
  \begin{tabular}{c|c|c|c|c|c} \hline
  \multirow{2}{*}{baseline-target}  & \multicolumn{2}{c|}{$\Delta$} & false alarms & \multicolumn{2}{c}{loss} \\ \cline{2-6}
  &  MW & Seq &  MW & MW & Seq \\ \hline
  animals-plants &  11 & 20 & 0 & -0.286 & -0.463  \\
  animals-fruit &  22 & 17 & 0 & -0.478 & -0.427 \\
  animals-fungus &  21 & 20 & 0 & -0.471 & -0.463 \\
  animals-fabric &  22 & 20 & 0 & -0.478 & -0.463 \\
  animals-garment &  5 & 16 & 5 & -2.582& -0.411 \\
  animals-music &  15 & 16 & 0 &  -0.391 & -0.411\\
  animals-weapon &  17 & 18 & 0 & -0.427 & -0.442 \\
  animals-tool &  4 & 18 & 0 & -0.054 & -0.442 \\
  plants-animals &  14 & 15 & 0 & -0.369 &-0.391 \\
  plants-fruit & 23 & 23 & 0 &  -0.483 & -0.483 \\
  plants-fungus & 1 & 19 & 0 & 0 & -0.453 \\
  music-tool &  9 & 23 & 1 & -0.719 & -0.483 \\
  music-weapon &  22 & 26 & 0 & -0.478 & -0.493 \\
  music-fabric &  18 & 20 & 0 & -0.442 & -0.463 \\
  music-garment &  10 & 23 &0 & -0.253 & -0.483 \\
   \hline
  \end{tabular}
%
\vspace{5mm}
\begin{tabular}{c|c|c|c|c|c} \hline
  \multirow{2}{*}{baseline-target}  & \multicolumn{2}{c|}{$\Delta$} & false alarms & \multicolumn{2}{c}{loss} \\ \cline{2-6}
  & MW & Seq  & MW & MW & Seq \\ \hline
  animals-plants & 4 & 16 & 6 & -3.054 & -0.411 \\
  animals-fruit & 14 & 16 & 0  & -0.369 & -0.411 \\
  animals-fungus & 16 & 19 & 0 & -0.411 & -0.453 \\
  animals-fabric & 13 & 17 & 3 & -1.844 & -0.427 \\
  animals-garment & 16 & 14 & 6 & -3.411 & -0.369 \\
  animals-music  & 10 & 15 & 0 &  -0.253 & -0.391 \\
  animals-weapon & 14 & 15 & 0  & -0.369 & -0.391 \\
  animals-tool & 16 & 13 & 6 & -3.411 & -0.344 \\
  plants-animals & 3 & 14 & 1 & -0.530 & -0.369 \\
  plants-fruit &  18 & 19 & 0  &  -0.442 & -0.453 \\
  plants-fungus & 1 & 18 & 0 & 0 & -0.442 \\
  music-tool  & 1 & 24 & 10 & -5.000 & -0.487 \\
  music-weapon & 21 & 29 & 0 & -0.471 & -0.497 \\
  music-fabric & 18 & 20 & 0 & -0.442 & -0.463 \\
  music-garment & 24 & 22 & 0 & -0.487 & -0.478 \\
   \hline
  \end{tabular}
  \caption{\label{tab.experiments.linear}
  Evaluation of MW ($\gamma=1$) and Sequential test based detection algorithm on 15 drift scenarios for the {\em linear drift} case for {\em KL} (top) and {\em cosine distance} divergence.  In all cases, both algorithms succeeded in detecting drift and {\bf Sequential never generated any false alarms}.\\
  \textbf{KLD (top)}: compared to Sequential, MW has a lower average detection delay (14.267 vs. 19.6)  but higher average false alarm rate (0.089 vs. 0). The overall average loss with MW is higher compared to Sequential (-0.527 vs. -0.451).\\
  \textbf{cosine (bottom)}: compared to Sequential, MW has a lower average detection delay (12.6 vs. 18.067) and higher average false alarm rate (0.315 vs 0). The overall average loss with MW is higher compared to Sequential (-1.366 vs. -0.426). } 
\end{table}
\section{Conclusion}\label{sec:discussion}

In our earlier work \citep{farchi_dube20}, we devised an efficient way to measure
changes in network performance at deployment time due to underlying change in the data observed, by extracting feature vectors and measuring their divergence over time.  Here, we extended on these experiments by using a framework for sequential control of the probability of false drift detection.  We also introduced a novel loss function which compares detection algorithms based on their detection delay and false alarm rate, adjusting for the amount of drift class contamination.  Overall, our results show that when compared with the non-sequential Mann-Whitney technique in our earlier work, sequential techniques can achieve much better false alarm control but with a slightly longer delay.  Our loss function can be customized to reflect the relative importance of these objectives to the user.





\vskip 0.2in
\bibliography{bibliography}


\newpage
\section{Appendix} \label{sec:appendix}

\subsection{Non-sequential Mann-Whitney test}
\label{appendix_mw}

\SetArgSty{textnormal}
\begin{algorithm}
\KwResult{DRIFT detected}
//Set your significance level \\
 $\alpha  = 0.05$;\\
 //Set observation window size \\  
  $\beta=40$;\\
  $t=\beta$;\\
   $t_d=\infty$;\\
 //Number of successive detections until decide drift\\
  $\gamma=1$;\\
  \text{count}=0;\\
  //Observe divergences\\
   $x_t=\mathcal{D}_{T,t;n}$;\\
  DRIFT = \textbf{False};\\
 \While{\text{DRIFT} = \textbf{False}}{
    $D_1=\{x_{t-\beta+1},\dots,x_{t-\beta/2}\};$\\
    $D_2=\{x_{t-\beta/2+1},\dots,x_t\}$;\\
    $P = \text{MannWhitney}(D_1,D_2)$ p-value;\\
    \If{$P < \alpha$}{
        //Detect drift\\
        \text{count}++;\\
        \uIf{\text{count}=$\gamma$}{
            $t_d=t$;\\
            DRIFT = \textbf{True};\\
        }
    }
    \Else{
            //Reset count and proceed\\
             \text{count}=0;\\
    }

    $t$++;
}
\Return $t_d$ (detection time)
 \caption{Non-sequential Mann-Whitney drift detection}
 \label{alg:mw}
\end{algorithm}

\subsection{Sequential testing}
\label{appendix_sequential_testing}
\subsubsection{Illustration of naive test with overlapping windows}
\label{sssec:peeking_problem}
Imagine we will observe $N=100$ data points in order over time; drift is said to occur if the sample mean $\bar{x}$ indicates the distribution mean $\mu$ is not equal to 0, as we expected it would.  We can consider the single hypothesis $H_0\colon \mu=0$ vs $H_A\colon \mu\ne0$. We can run a Student's T-test to see if the mean of an observed sample indicates $H_0$ is false, that is, that drift has occurred.  The tendency of such a test to produce false alarms can be evaluated by simulating this $N=100$ under the condition of non-drift, and seeing how often the test (falsely) rejects $H_0$.

Say we draw $r=10,000$ independent and identically-distributed (iid) samples $\{x_1,\dots,x_{100}\}$ from the standard normal distribution $\text{N}(0,1)$.  These are drawn under the mean value $\mu=0$ specified in the null hypothesis.  What can we say about the false alarm probability?

\begin{itemize}
    \item \textbf{Single hypothesis test for each sample of size $N=100$}: In this case, on average in $10,000\times\alpha$ samples, the sample mean will be unusually different enough from 0 so as to falsely reject the $H_0\colon \mu=0$, even though the sample was drawn from the distribution.
    \item \textbf{Multiple hypotheses on non-overlapping windows}: Say we partition each of the $r$ samples into $w=5$ non-overlapping subsets (e.g., 1--20, 21--40,\dots,81--100). The T-test is applied on each subset separately, each of which again falsely rejects $H_0$ with probability $\alpha$.  Since the subsets are non-overlapping, the $w=5$ tests are independent.  Thus, if we now declare drift on one of the 100-size samples if \textit{any of the 5 subsets reject $H_0$ and declare drift}, the probability we will make the correct decision of not declaring drift is now $(1-\alpha)^w=(1-0.05)^5\approx 0.77$, instead of $1-\alpha=0.95$ with a single test, even though $\alpha$ did not change.  This shows the need to adjust our decision procedure if we wish to control the false alarm probability but also conduct multiple tests (i.e., `peek' at the data).  This area of study is called `multiple hypothesis testing'.
\end{itemize}

Each test applied above was independent of the others (since the samples didn't overlap) and also typically requires a minimum sample size (i.e., not a single observation). Say we now wanted to `peek' at the data as it was being received, as in the second instance, but rather than only observing a subset, we want to test based on the entire history observed up to each observation.  That is, say we conduct the test on each of the nested 81 subsets $\{x_1,\dots, x_j\}$ for $j=20,\dots,100$, and declare drift has happened on the first subset to reject $H_0$ (its p-value $<\alpha$).  In reality, we would stop observing after the first declaration (we have already made a decision), but assume we continue testing until $x_{100}$.  Let $V$ be the number of such p-values out of 81 that are $<\alpha$.  A false alarm occurs on each of the $r$ samples if its $V\geq0$.  The false alarm probability is best analyzed by simulation and considering the observed $\text{Pr}(V\geq 0)$ over the 10,000 draws.  Results from one experiment are shown below, for various thresholds $\alpha$:

\begin{table}[H]
\centering
\begin{tabular}{r|rrrr}
   \hline
$\alpha$ & 0.05 & 0.01 & 0.005 & 0.001 \\ 
\hline
  $\text{Pr}(V\geq 1)$ & \bf{0.2296} & 0.0678 & 0.0360 & 0.0074 \\ 
  $\text{E}(V)$ & 4.0825 & 0.7680 & 0.3598 & 0.0527 \\ 
   \hline
\end{tabular}
\end{table}

Thus, for instance, if $\alpha=0.05$, we would in fact make a false alarm in approximately 23\% of these draws, and we would make around 4 false declarations if we observed all the 100 observations each time.  In reality, this is the type of test we would want to conduct for drift detection, but the actual false alarm rate is unacceptably higher than the desired 5\%.

\subsubsection{Change Point Models (CPM) details}
\label{subsec:cpm_details}
\label{appendix_cpm}

\SetArgSty{textnormal}
\begin{algorithm}[h]
\SetAlgoLined
\KwResult{DRIFT detected}
 //Set your significance level \\
  $\alpha  = 0.05$;\\
 //Initial stabilization period\\
  $b=25$;\\
  $t=b$; \\
  $t_s=\infty;\: K=\infty$;\\
 //Sequential critical values\\
 $h_b,\:h_{b+1},\dots$\\
 DRIFT = \textbf{False}

 \While{\text{DRIFT} = \textbf{False}}{
    //Consider each possible split of data into two before/after subsets\\
    \For{$k=2,\dots,t-1$}{
        $\textbf{X}_0=\{x_1,\dots,x_k\}$\\
        $\textbf{X}_1=\{x_{k+1},\dots,x_t\}$\\
        $W_{k,t} = \text{norm}(\text{Diff}(\textbf{X}_0,\textbf{X}_1))$\\
    }
    //Find most significant split\\
    $\hat{\tau}=\text{argmax}_k\: W_{k,t}$\\
    \If{$W_{\hat{\tau},t} > h_t$}{
       $t_d=t;\: K=\hat{\tau}$;\\
       DRIFT = \textbf{True}
    }
    $t$++;
    
    }
    \Return $t_d$ (detection time), $K$ (changepoint, $:=t_s-1$); both are $\infty$ if no detection
 \caption{Outline of Ross \& Adams' CPM algorithms}
 \label{alg:cpm}
\end{algorithm}

\subsection{Step drift}
\label{subsec:step_drift}
\subsubsection{Divergence distribution}
\begin{figure*}[htb]
\begin{multicols}{2}
    \includegraphics[width=\linewidth]{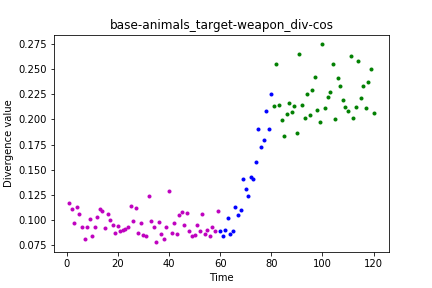}\par 
    \includegraphics[width=\linewidth]{figures/linear/cos/animals_fungus.png}\par
    \end{multicols}
 \begin{multicols}{2}
     \includegraphics[width=\linewidth]{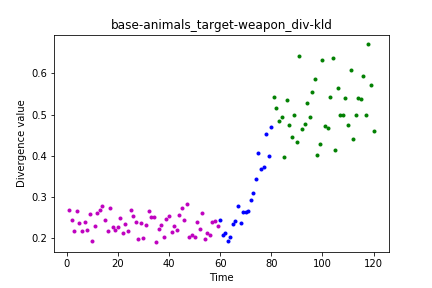}\par
     \includegraphics[width=\linewidth]{figures/linear/kld/animals_fungus.png}\par
 \end{multicols}
 \caption{{\bf Linear drift:} Divergence distribution over the experiment interval for 2 example scenarios using cosine distance (top) and KLD (bottom).  Drift starts at $t=60$ and ends at $t=80$. The value of divergence is close to 0 for $t < t_s$, has an increasing trend for $t_s \leq t \leq t_e$ and remains high for $t > t_e$. \label{fig:div-distribution-linear-1}}
\end{figure*}
\begin{figure*}[htb]
\begin{multicols}{2}
    \includegraphics[width=\linewidth]{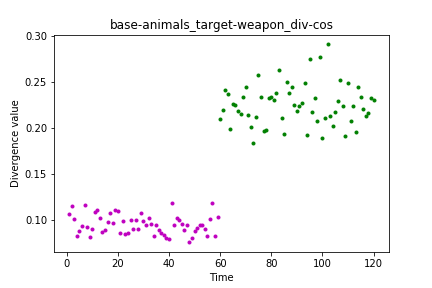}\par 
    \includegraphics[width=\linewidth]{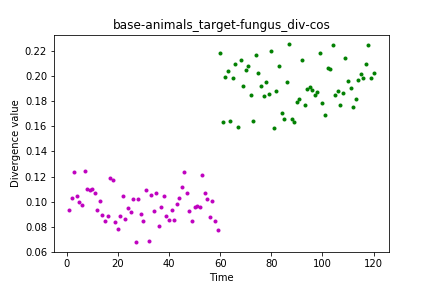}\par
    \end{multicols}
\begin{multicols}{2}
    \includegraphics[width=\linewidth]{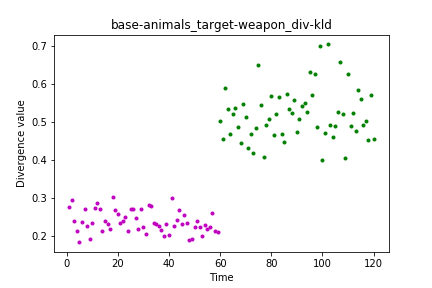}\par
    \includegraphics[width=\linewidth]{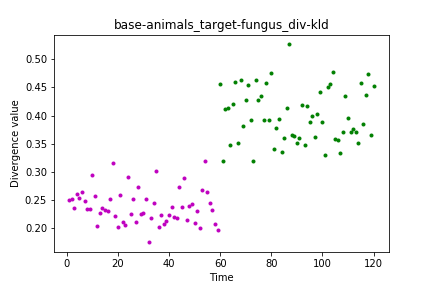}\par
\end{multicols}
\caption{{\bf Step drift:} Divergence distribution over the experiment interval for 2 example scenarios using cosine distance (top) and KLD (bottom) for with $t_s=t_e=60$. The value of divergence is close to 0 for $t < t_s$, and high for $t \geq t_s$. \label{fig:div-distribution-step}}
\end{figure*}

Compared to linear drift (see Figure~\ref{fig:div-distribution-linear-1}), in step drift the divergence distribution has an abrupt switch at $t=t_s$ as see in Figure~\ref{fig:div-distribution-step}. Thus for $t<t_s$, divergence values are very low and for $t\geq t_s$ the divergence values are all high. 

\subsubsection{Change detection}
The comparison between MW and Sequential for step drift scenario is shown in Table~\ref{tab.experiments.step}. Compared to linear drift case, the detection delay here is much smaller since the drift is sudden and thus easier to detect.  Note that the loss function (Section~\ref{sec:loss_function}) adjusts the penalization of detection delay for the drift contamination $\textbf{p}$, thus results can be compared across different scenarios.  Although it depends on the values chosen for the two penalties $C_1$ and $C_2$, the loss for the step drift case seems to be larger than the linear drift because even with sudden drift, some delay is typically required for the statistical difference between pre- and post-drift to become statistically significant.

\begin{table*}[t]
\centering
\hspace{-0.2in}
  \begin{tabular}{c|c|c|c|c|c|c|c|c|c|c|c} \hline
  \multirow{2}{*}{baseline} & \multirow{2}{*}{target} & \multicolumn{2}{c|}{detection} & \multicolumn{2}{c|}{$\Delta$} & \multicolumn{2}{c|}{false alarms} & \multicolumn{2}{c|}{loss} & \multicolumn{2}{c}{$\theta$}\\ \cline{3-12}
  & & MW & Seq & MW & Seq & MW & Seq & MW & Seq & MW & Seq\\ \hline
  animals & plants & Yes & Yes& 6 & 7 & 0 & 0 & -0.412 & -0.438 & 0 & 0\\
  animals & fruit & Yes & Yes& 8 & 7 & 3 &0 & -1.956 & -0.438 & 0.75 & 0 \\
  animals & fungus & Yes & Yes& 4 & 7 & 2 & 0 & -1.323 & -0.438 & 0.667 & 0 \\
  animals & fabric & Yes & Yes& 9 & 6 & 5 & 0 & -2.969 & -0.412 & 0.833 & 0 \\
  animals & garment & Yes & Yes& 5 & 6 & 0 & 0 & -0.375 & -0.412 & 0 & 0\\
  animals & music & Yes & Yes & 7 & 7 & 0 & 0 &  -0.438 & -0.438 & 0 & 0\\
  animals & weapon & Yes & Yes & 10 & 7 & 7 & 0 & -3.978 & -0.438 & 0.875 & 0 \\
  animals & tool & Yes & Yes& 1 & 7 & 8 & 0 & -4.000 & -0.438 & 0.889 & 0\\
  plants & animals & Yes & Yes& 4 & 7 & 14 & 0 & -7.323 &-0.438 & 0.933 & 0\\
  plants & fruit & Yes & Yes& 6 & 7 & 0 & 0 &  -0.412 & -0.438& 0 & 0 \\
  plants & fungus & Yes & Yes& 8 & 7 & 0 & 0 & -0.456 & -0.438 & 0 & 0 \\
  music & tool & Yes & Yes & 6 & 7 & 1 & 0 & -0.912 & -0.438 &  0.5 & 0\\
  music & weapon & Yes & Yes& 6 & 8 & 0 & 0 & -0.412 & -0.456 & 0 & 0\\
  music & fabric & Yes & Yes& 5 & 7 & 0 & 0 & -0.375 & -0.438 & 0 & 0\\
  music & garment & Yes & Yes& 1 & 7 & 3  & 0 & -1.500 & -0.438 & 0.75 & 0\\
   \hline
  \end{tabular}

\vspace{5mm}

 \begin{tabular}{c|c|c|c|c|c|c|c|c|c|c|c} \hline
  \multirow{2}{*}{baseline} & \multirow{2}{*}{target} & \multicolumn{2}{c|}{detection} & \multicolumn{2}{c|}{$\Delta$} & \multicolumn{2}{c|}{false alarms} & \multicolumn{2}{c|}{loss} & \multicolumn{2}{c}{$\theta$}\\ \cline{3-12}
  & & MW & Seq & MW & Seq & MW & Seq & MW & Seq & MW & Seq\\ \hline
  animals & plants & Yes & Yes& 6 & 6 & 0 & 0 & -0.412 & -0.412 & 0 & 0\\
  animals & fruit & Yes & Yes& 9 & 7 & 2 & 0 & -1.469 & -0.438 & 0.667 & 0 \\
  animals & fungus & Yes & Yes& 2 & 7 & 4 & 0 & -2.146 & -0.438 & 0.8 & 0 \\
  animals & fabric & Yes & Yes& 1 & 7 & 7 & 0 & -3.5 & -0.438 & 0.875 & 0 \\
  animals & garment & Yes & Yes& 7 & 7 & 2 & 0 & -1.438 & -0.438 & 0.667 & 0\\
  animals & music & Yes & Yes & 9 & 7 & 0 & 0 &  -0.469 & -0.438  & 0 & 0\\
  animals & weapon & Yes & Yes & 8 & 7 & 8 & 0 & -4.456 & -0.438 & 0.889 & 0 \\
  animals & tool & Yes & Yes& 8 & 7 & 0 & 0 & -0.456 & -0.438 & 0 & 0\\
  plants & animals & Yes & Yes& 4 & 7 & 3 & 0 & -1.823 & -0.438 & 0.75 & 0\\
  plants & fruit & Yes & Yes&  3 & 7 & 0 & 0 &  -0.25 & -0.438 & 0 & 0 \\
  plants & fungus & Yes & Yes& 10 & 7 & 5 & 0 & -2.978 & -0.438 & 0.833 & 0 \\
  music & tool & Yes & Yes & 6 & 7 & 2 & 0 & -1.412 & -0.438 & 0.667 & 0\\
  music & weapon & Yes & Yes& 5 & 8 & 1 & 0 & -0.875 & -0.456 & 0.5 & 0\\
  music & fabric & Yes & Yes& 6 & 7 & 0 & 0 & -0.412 & -0.438 & 0 & 0\\
  music & garment & Yes & Yes& 1 & 7 & 3  & 0 & -1.5 & -0.438 & 0.75 & 0\\
   \hline
  \end{tabular}
    \caption{\label{tab.experiments.step} Evaluation of MW ($\gamma=1$) and Sequential test based detection algorithm on 15 drift scenarios for the {\em step drift} case for {\em KL} (top) and {\em cosine distance} divergence cases.   In all cases, both algorithms succeeded in detecting drift and Sequential never generated any false alarms.\\
    \textbf{KLD (top)}: compared to Sequential, MW has a lower average detection delay (5.733 vs. 6.933)  but higher average false alarm rate (0.413 vs. 0). The overall average loss with MW is higher compared to Sequential (-1.789 vs. -0.435).\\
    \textbf{cosine (top)}: compared to Sequential, MW has a lower average detection delay (5.667 vs. 7) but higher average false alarm rate (0.493 vs 0). The overall average loss with MW is higher compared to Sequential (-1.573 vs. -0.437). }
\end{table*}

\end{document}